**Accepted for publication in NANO LETTERS**

Compact antenna for efficient and unidirectional launching and decoupling of surface plasmons

Alexandre Baron[1], Eloïse Devaux[2], Jean-Claude Rodier[1], Jean-Paul Hugonin[1], Emmanuel Rousseau[1#], Cyriaque Genet[2], Thomas W. Ebbesen[2] and Philippe Lalanne[1,3]

[1]Laboratoire Charles Fabry, Institut d'Optique, Univ Paris-Sud, CNRS, Campus Polytechnique RD 128, 91127 Palaiseau cedex, France.

[2]ISIS, Université de Strasbourg et CNRS (UMR 7006), 8 allée Monge, 67000 Strasbourg, France.

[3]Laboratoire Photonique, Numérique et Nanosciences, Institut d'Optique, Univ Bordeaux 1, CNRS, 33405 Talence cedex, France.

[#]Present address : Université Montpellier 2, Laboratoire Charles Coulomb UMR 5221, CNRS, F-34095, Montpellier, France

Keywords : surface plasmon polariton, unidirectional launcher and decoupler, plasmonic nanoantenna, diffraction gratings.

ABSTRACT :

Controlling the launching efficiencies and the directionality of surface plasmon polaritons (SPPs) and their decoupling to freely propagating light is a major goal for the development of plasmonic devices and systems. Here, we report on the design and experimental observation of a highly efficient unidirectional surface plasmon launcher composed of eleven subwavelength grooves, each with a distinct depth and width. Our observations show that, under normal illumination by a focused Gaussian beam, unidirectional SPP launching with an efficiency of at least 52% is achieved experimentally with a compact device of total length smaller than 8 μm. Reciprocally, we report that the same device can efficiently convert SPPs into a highly directive light beam emanating perpendicularly to the sample.

**Introduction.** As the demand for increasing speed and bandwidth of information processing rises, plasmonic devices and circuits present an interesting approach, since they represent an enabling technology that bridges electronic and photonic physical phenomena on similar sub-micron scales. So far a large diversity of active and passive optical nanodevices based on surface-plasmon-polaritons (SPPs) have been theoretically proposed and experimentally demonstrated [1-5]. Unfortunately, many problems remain open and need to be solved before achieving the full potential of surface plasmon technology. In that respect an important issue is to efficiently couple SPPs from freely-propagating light and to control the direction in which they are launched. Conventional SPP couplers involve prisms or subwavelength gratings [6], and rely on interaction lengths that are comparable to the SPP damping characteristic lengths. In view of potential applications, it is essential to miniaturize the devices and to reinforce the SPP interaction strength [7-9], so that a unidirectional launch pad for electromagnetic surface waves may provide the missing link between conventional optics and future highly integrated photonic chips.

Two important figure-of-merits of SPP launchers are the launching efficiency $\eta$ in the desired direction and the extinction ratio $E_R$ defined as the ratio between the SPP intensity launched into the desired direction and the intensity launched into the opposite direction. Recently, Delacour *et al.* [10] built a device that exhibits a silicon/slot-copper-line/silicon transmission of 40% to convert guided photons into plasmons and back to guided photons. And in 2010, Briggs *et al*. [11] achieved an impressive 80% coupling from a semiconductor nanowaveguide to a polymer-on-gold dielectric-loaded waveguide. However these two approaches rely on integrated devices that use confined fields. The manufacture of an efficient unidirectional launcher from freely-propagating light has received considerable attention over the past few years. For instance, Choi *et al*. [12] have fabricated an asymmetric Bragg resonator. Under illumination by a femtosecond temporal-phase-controlled pulse, an extinction ratio of 10 was achieved. Wang *et al.* [6] designed a tunable unidirectional launching device that consists in two silver films with a nanoslit on the top film; an extinction ratio of 20 was theoretically predicted, but to our knowledge the device has not yet been fabricated. By illuminating a slit located next to a SPP Bragg reflector composed of several periodic grooves, unidirectional launching has been observed [13]; however the launching efficiency (not reported) is supposedly rather small since light is coupled from the rear side to the subwavelength slit, first. More recently, Chen *et al.* [14] reported a remarkable ($E_R$ = 30) unidirectional SPP generation using an asymmetric single-nanoslit, but again no information on the coupling efficiency was provided.

In this letter, we demonstrate a novel device composed of eleven grooves. In our opinion, the device offers better performance than those reported so far [1-2,6-10,13,14] with an experimental unidirectional launching efficiency η larger than 52% and a large extinction ratio $E_R = 47$ at λ = 800 nm. In addition, the device is compact (8-μm long) and operates under normal illumination by a Gaussian laser beam that is conveniently impinging at normal incidence. It is also shown that the same device, illuminated by an SPP, can efficiently convert the SPP into a freely propagating beam that escapes the sample normally to the surface with a divergence angle smaller than 15°. The experimental results are all supported by fully-vectorial computational data obtained with a frequency-domain aperiodic-Fourier Modal method (a-FMM) [15,16].

**Design.** The design of the unidirectional SPP launcher is performed with the a-FMM at λ = 800 nm. Before optimization, we consider a device composed of eleven identical grooves periodically placed on a gold interface illuminated by a Gaussian beam polarized perpendicularly to the grooves (TM polarization) with a waist equal to 4.8 μm, see Fig. 1(a). We assume a finite metal conductivity with a gold refractive index 0.18+5.13i, taken from the tabulated data in [17]. The electromagnetic field scattered by the device is first calculated with the a-FMM [15, 16]. Then by computing an overlap integral between the SPP field of the flat interface (known analytically) and the electromagnetic field scattered on the left- and right-hand sides of the devices, the SPP excitation coefficients, $\alpha^+(x > x_0)$ and $\alpha^-(x < -x_0)$ with $x_0$ large enough, on both sides of the device are determined. Details of the orthogonality relation used to calculate the overlap integral and to extract the SPP excitation coefficients from the near-field pattern can be found in [18]. After normalizing the coefficients by the incident power of the Gaussian beam, the launching efficiencies, $\eta^+ = |\alpha^+(x > x_0)|^2$ and $\eta^- = |\alpha^-(x < -x_0)|^2$, are determined. This procedure is then repeated several times, varying the depths, widths and locations of the eleven grooves independently. For the optimization, we use the simplex search Nelder-Mead method. This direct method, that does not use numerical or analytic gradients, relies on an iterative simplex-minimization approach that progressively reduces the explored volume in the parameter hyper-space [19].

The cost value used for the optimization is simply the efficiency $\eta^+$ of the SPP launched on the right side of the device. We have also tried cost functions such as $1+\eta^+-\eta^-$ to favor geometries with a large extinction ratio, but somehow this resulted in smaller $\eta^+$ values. The optimization procedure is repeated several times by starting with different initial groove widths and depths, ranging from 50 to 200 nm. We observe that many geometries that look different may provide large efficiencies ($\eta^+ > 50\%$) and large extinction ratios ($E_R \sim 30$-$50$). One of our best optimized geometry is shown in Fig. 1b. Starting from the left side, the eleven groove depths and widths are 102, 92, 70, 65, 55, 72, 77, 58, 56, 50, 40 nm and 170, 157, 190, 270, 320, 424, 422, 435, 419, 428, 443 nm, respectively. The corresponding groove-center locations are -3.65, -2.90, -2.21, -1.50, -0.79, 0, 0.73, 1.48, 2.26, 3.02 and 3.76 μm. For these parameters, the SPP efficiency $\eta^+$ is as large as 62% ($\eta^- = 1\%$), implying that two thirds of the incident photons are converted into a single SPP launched on the right side of the device.

The design procedure relies on a global approach in which the electromagnetic field scattered by the device is repeatedly optimized to maximize the total SPP launching efficiency. Indeed, at a microscopic level, the collective optical response of the 11-groove launcher is governed by two distinct elementary waves that are initially launched on the surface by every individual groove and then propagate on the surface, before further interacting with nearby grooves. The elementary waves contain two distinct fields, a SPP bounded to the surface and a quasi-cylindrical wave formed by radiative and evanescent components that persists along the surface over a propagation distance of a couple of wavelengths [20]. In principle, the physics of the unidirectional launcher could have been revealed by calculating the elementary scattering coefficients of every individual groove and by writing SPP-coupled mode equations to analytically derive the launcher efficiencies [21-23].

In order to get an insight into the multiple interference process responsible for the efficient unidirectional launching, we have calculated the excitation rates of the SPPs launched on every flat part (the ridges) of the surface in between the grooves. The excitation rates are shown in Fig. 1(b) with the superimposed white and green curves. The latter represent the modulus squared of the SPP scattering coefficients $\alpha^+(x)$ and $\alpha^-(x)$ calculated with the overlap integral. It is remarkable to observe that, as one moves from the left to the right sides, $|\alpha^+(x)|^2$ keeps on increasing to reach a value of 62% for $x > x_0$ and that, except for the first five ridges on the left for which $|\alpha^{-}(x)|^2 \approx |\alpha^{+}(x)|^2$, $|\alpha^{+}(x)|^2$ is much larger than $|\alpha^{-}(x)|^2$. We additionally note that $|\alpha^{-}(x)|^2$ reaches its maximum value (15%) in the central part of the device on the fifth ridge. Finally, it should be emphasized that $|\alpha^{-}(x)|^2$ and $|\alpha^{+}(x)|^2$ not only take into account the multiple scattering of the different SPPs that are launched and scattered by every individual groove, but also consider the cross-conversion from quasi-CW to SPPs through scattering at adjacent grooves [21-23]. As such, they should be considered as an exact measure of the SPP excitation strength inside the device.

Then we evaluate the performance of the launcher as an SPP decoupler. For that purpose, we calculate the field scattered by the device when it is illuminated by an SPP impinging from the right side. The magnitude of the magnetic field is shown in Fig. 1(c). Actually most of the energy is scattered out in free space; the SPP transmittance is almost null (<1%), the SPP reflectance is close to 1% and the device absorption remains reasonably weak (22%). By performing a plane wave decomposition of the scattered field, we find that the scattered light is highly directive: actually, as much as 75% of the incident SPP energy is radiated out into a cone with a ±10° extraction angle around the surface normal, see the inset in Fig. 1(c). Not only does the designed device act as an efficient unidirectional launcher, but it also acts as an SPP-to-free-space converter that radiates a well-shaped directive beam emanating perpendicularly to the sample.

**Fabrication.** Encouraged by these theoretical predictions, we decided to fabricate and test the actual performance in real situation. Figure 2a shows the 3-component device that we use to test the performance of the component acting either as an SPP launcher or an SPP decoupler. The device is assumed to be illuminated by a Gaussian beam focused onto the central component. The latter launches SPPs on its right- and left-hand sides with efficiencies $\eta_c^{+}$ and $\eta_c^{-}$, respectively ($\eta_c^{+} >> \eta_c^{-}$ according to the design). At a distance *d* from the launcher (see Fig. 2(a) for a strict definition of *d*), the same device is used to scatter the launched SPPs out of the sample with an efficiency $\eta_d$. Note that the right-hand side decoupler is flipped so that the SPP propagating to the right meets the shallower groove first.

The subwavelength grooves are etched in a 300-nm-thick gold film sputtered on a glass substrate. The grooves are prepared with a Dual Beam Strata 235 Focused Ion Beam. According to the designed values, the separation distances and widths of the grooves are milled with a precision of +/- 10 nm, measured with a scanning electron microscope (SEM), see the images in Figs. 2(b1)-2(b4); the groove depths are harder to control. Subsequent atomic force microscopy (AFM) characterization led us to a precision of +/- 25 nm for the deeper grooves to +/- 5 nm for the shallower ones. Although the designed depths were not all perfectly matched in the initial sample tests, we decided to fabricate the three-component device to perform optical measurements as the component appeared to be efficient in launching SPPs. SEM pictures of the fabricated sample are shown in Fig. 2(b). Figure 2(b2) shows the whole device composed of the launcher in the center and of two decouplers situated on the right and left hand-sides. Each component is 7.9-μm-wide. On the same sample, an array of 3×3 devices are fabricated. The array is composed of three subsets, each containing three supposedly identical devices with the same separation distance *d* between the launchers and the decouplers. Thus a total number of nine devices with three different separation distances, $d = 12.2$, 22.2 and 32.2 μm, is available.

**Characterization.** To characterize the sample, we use the setup sketched in Fig. 3(a). A broadband supercontinuum white light source connected to an acousto-optic tunable filter delivers monochromatic CW-light (2-nm FWHM linewidth) in the 650-1000 nm spectral range. The output light wavelength is set to $\lambda$ = 800 nm for the experiment. It is spatially filtered by a photonic-crystal fiber (PCF) and collimated using a microscope objective (Ob1: ×28). The TM polarization is controlled via the polarizer (P1). The beam is then focused on the sample using a second microscope objective (Ob2: ×16) with a numerical aperture of 0.35. The sample is mounted on an XYZ-translation stage. The light scattered back from the sample is collimated by the same objective Ob2 and is extracted from the main beam with a beam splitter (BS1). It is subsequently focused on a CCD camera using an AR-coated 500-mm-focal lens (L). In order to ease the alignment procedure, a red light-emitting-device additionally illuminates the sample through a second beam splitter (BS2). The doublet (L-Ob2) provides a magnification ×40. The objectives Ob1 and Ob2 are chosen so that the waist of the incident Gaussian beam on the sample is $w_0 \sim 6.5\ \lambda$, a value close to the 6λ-waist used for the design. The beam waist is measured by acquiring an image of the light reflected on the flat gold surface of the sample. The spot in the image is then fitted by a Gaussian function. Note that the chromatic aberration of the (L-ob2) doublet is negligible; it results in a slight defocusing distance smaller than 5 μm and much smaller than the Rayleigh length $z_R = \pi w_0^2/\lambda \sim 100$ μm of the Gaussian beam.

To accurately measure the power backscattered from the sample, the images (Fig. 3(b)-(c)) obtained with the CCD are further processed. First a square integration window of area $a\times a$ is chosen around the position of interest (Fig. 3(c)). Then the raw power $P_R$ received by the CCD in the integration is calculated numerically by adding the pixel intensities. In Fig. 3(d), the black curves show $P_R$ as a function of the size of the integration window. Then, the same window is positioned away from the illuminated area to measure the background noise $P_N$, see the blue curves. As shown by the red curve, the difference $P_R-P_N$ always saturates for large enough integration windows, and the saturation value $P^+ = P_R-P_N$ (Fig. 3(d)) obtained for $a\to\infty$ represents the actual scattered power. The same procedure is used to measure $P_0$, by shining light directly on the gold surface away from any device. For all our measurements, we check that a flat saturation is reached.

To measure the total coupling/decoupling efficiencies, $\eta_c^+\eta_d$ and $\eta_c^-\eta_d$, the Gaussian beam is focused on the central launcher component. The horizontal and vertical positioning of the sample is optimized so as to maximize the power $P^+$ collected on the component located on the right-hand side of the device. Systematically, we observe that $P^+$ is maximal when the infrared spot is centered on the device, see Fig. 3(c). Prior to each measurement, an image of the incident beam reflected on a flat part of the gold surface is taken in order to check the beam waist and incident power stabilities. For a gold refractive index of 0.18+5.13i at $\lambda = 800$ nm [16], the reflection coefficient is $R = 97\%$. Thus $P_0/R$ represents the incident power on the launcher. The total coupling/decoupling efficiencies for the right- and left-propagating SPPs can be respectively written as $\eta^+ = \eta_c^+\eta_d \exp(-d/l_e) = RP^+/P_0$ and $\eta^- = \eta_c^-\eta_d \exp(-d/l_e) = RP^-/P_0$, where the exponential terms take into account the damping of the launched SPPs over the separation distance $d$. $l_e$ is the SPP extinction length [24].

The $\eta^+$ and $\eta^-$ efficiencies are measured for the nine devices. For every measurement, the XYZ-translation stage is monitored so as to maximize $\eta^+$. The nine values are shown with triangles, squares and circles in Fig. 4(a) as a function of $d$. Note that the measured efficiencies are remarkably similar, since the maximal relative error between the measured intensities is not greater than 1% for every $d$ = 12.2, 22.2 and 32.2 μm. In a log-scale, the solid line represents a linear fit, $ln(RP^+/P_0) = -d/l_e+ln(\eta_c^+\eta_d)$. The vertical-intercept of the linear fit gives the experimental value of the product $\eta_c^+\eta_d$ and the slope represents the extinction length, $l_e$ = 16±2 μm, a value that is three-fold smaller than the theoretical value of 45 μm, obtained from the classical SPP damping rate formula with a gold refractive index of 0.18+5.13i. We attribute this discrepancy to surface and bulk roughness of the thick sputtered film. This roughness can be seen on the close-up SEM pictures of the device: on the surface of the gold film (Fig. 2(b4)) and at the bottom of the grooves (Fig. 2(b3)). In the end, from the fit, we find $\eta_c^+\eta_d$ = 0.38 (and $\eta_c^-\eta_d$ = 0.008 similarly).

From the sole knowledge of the product $\eta_c^+\eta_d$, one should make a hypothesis to infer the launching and decoupling efficiencies. Hereafter, we assume that the experimental value of the decoupling efficiency is equal to the theoretical value. Additional calculations were provided to estimate the impact of inevitable fabrication errors. We found that a ±10% uniform error on all the groove depths degrades the coupling and decoupling efficiencies similarly by ~0.06, showing that the design is quite robust. Thus, we can reasonably expect that the experimental value of $\eta_d$ is less than its computed value of 75%, and that the inferred value of $\eta_c^+$ is likely to represent a lower-bound. With this assumption, the launching (for the right propagating SPP) is found to be at least equal to 52%, an experimental value larger to all those reported so far to our knowledge.

Next, by tuning the accousto-optic tunable filter, we vary the laser wavelength so as to measure $RP^+/P_0$ and $RP^-/P_0$ as a function of λ in order to obtain the spectroscopic behavior of the launching device. From these measurements, we infer the launching efficiency $\eta_c^+(\lambda)$ under the assumption that the experimental decoupling efficiency is equal to the theoretical efficiency $\eta_d$ shown with the long-dashed black curve in Fig. 4(b). As pointed out in the previous paragraph, the inferred values of $\eta_c^+(\lambda)$ are likely to represent lower bounds. The experimental data for $\eta_c^+(\lambda)$ plotted as red dots fairly agrees with the calculated data (red curve), showing that, in accordance with the numerical predictions, $\eta_c^+$ is maximum for λ = 800 nm and presents a ~ 35 nm half-width half-maximum response, which is reasonably close to the numerical value of 25 nm. Furthermore, we notice numerically that the decoupling efficiency remains large for λ > 800 nm. Essentially, as the wavelength increases, the nice diffraction lobe in the left inset of Fig. 1(c) first shifts towards negative θ and then splits into two main lobes that remain within the ±20° acceptance angle of the 0.35 NA microscope objective used for the experiment. By the way, note that the decoupling efficiency shown in Fig. 4(b) is defined as the normalized energy scattered out into a ±20° cone.

We then measure $RP^+/P_0$ and $RP^-/P_0$ as the horizontal translation stage of the sample is monitored. We denote by $\delta X = X - X_0$ the deviation from the position $X_0$ that maximizes the launching SPP efficiency toward the right. $\delta X$ is positive, when the device is shifted leftward in Fig. 3(b). According to our computational results (Fig. 4(c)), the launching of the right-propagating SPP is expected to be maximal for $\delta X = 0$, whereas the launching of the left-propagating SPP is expected to rise for negative values of δX. In our setup, $\delta X$ is measured directly on the CCD by recording an image with the incoherent red illumination. The accuracy is just limited by the pixel resolution of the CCD, and with 9.9×9.9 $\mu m^2$ pixels and a magnification factor of 40 for the imaging doublet, the absolute precision on $\delta X$ is ~100 nm. From the measurements of $RP^+/P_0$ and $RP^-/P_0$, we infer lower-bounds for the launching efficiencies $\eta_c^+(\delta X)$ and $\eta_c^-(\delta X)$, under the reasonable assumption that $\eta_d = 75\%$ is independent of $\delta X$. A series of CCD images recorded for various sample positions is shown in Fig. 4(d). We note that the decoupled light ($P^-$) on the left-hand side is generally much weaker than the decoupled light ($P^+$) on the right hand-side. As can be seen in Fig. 4(c), $\eta_c^+$ is maximum when the launcher is centered ($\delta X = 0$), whereas $\eta_c^-$ is maximum for $\delta X = -4$ µm. The experimental data shown with circles agrees well with the theoretical predictions obtained for the designed device. Let us note that for $\delta X > 0$, $P^-$ is too weak and cannot be accurately measured even by changing the integration time of the CCD.

**Discussion and Conclusion.** In addition to the launching efficiency, an important figure-of-merit is the so-called extinction ratio $E_R$, defined as the quotient between $\eta_c^+$ and $\eta_c^-$. For the present device, the extinction ratio defined for $\delta X = 0$ is measured to be $E_R = 47$ ($\eta_c^- = 1.1\%$ and $\eta_c^+ = 52\%$). The two figures-of-merit are remarkably high with respect to previous results [1-2,6-10,13,14]. In addition, they are achieved for normal incidence; this is really convenient in most applications. With eleven grooves, the device length is presently 8 µm, i.e. 10λ for operation at λ = 800 nm. Other calculations performed during the earlier stages of the design let us expect that even more compact launchers (6λ-long for instance) may be manufactured with similar launching performances. Note however that the decoupling process has not been explored for such devices. We expect that the present approach may be of interest for various SPP applications, including the manufacture of compact SPP sensors by replication techniques of a master in a polymer, followed by a metallization by sputtering for instance.

The authors acknowledge initial discussions with J. Gierak and E. Bourhis who were involved at the earlier stage of the project. They also thank V.-H. Mai, L. Jacubowiez, F. Marquier and J.J. Greffet for fruitful discussions and help, and Buntha Ea Kim and Anne-Lise Courtot for providing SEM pictures of the sample. The authors also thank the Laboratoire de Photonique et Nanostructures (Marcoussis) for pictures taken with a SEM Hitachi S4800. This research was supported in part by the ERC (Grant n°227577).

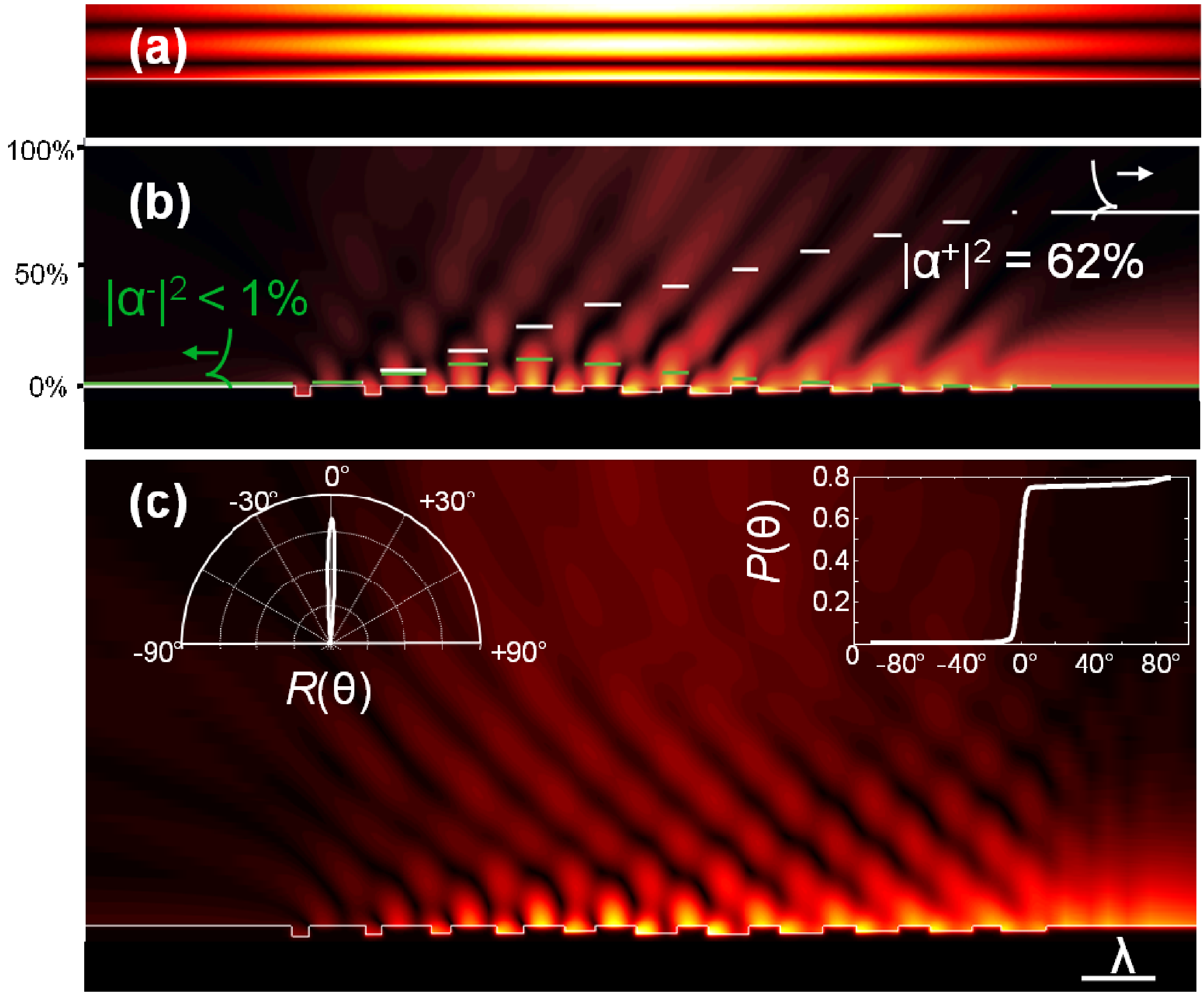

FIG 1

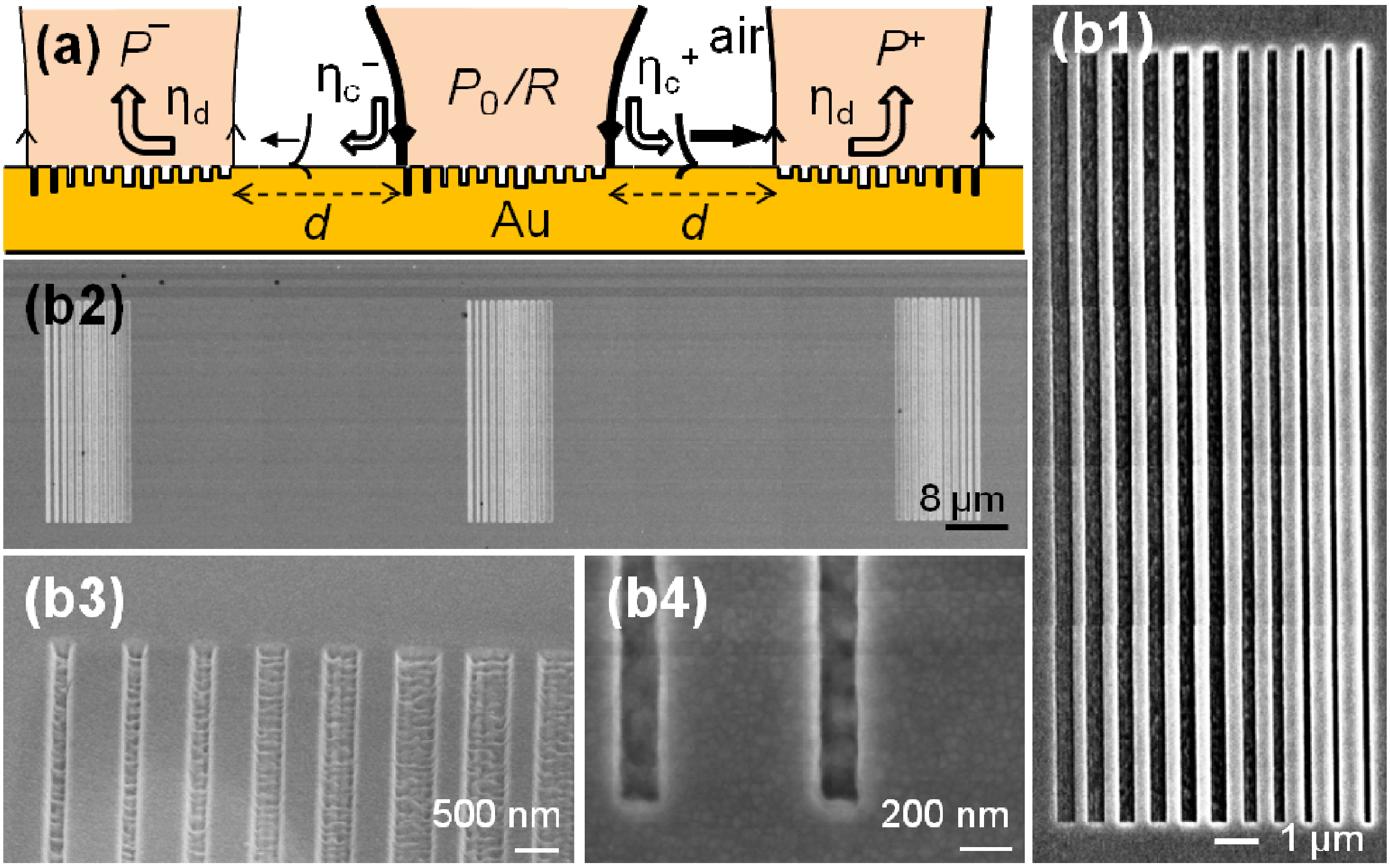
(a) $P^-$
$\eta_d$
$\eta_c^-$
$P_0/R$
$\eta_c^+$
air
$P^+$
$\eta_d$
$d$
Au
$d$
(b2)
8 µm
(b3)
500 nm
(b4)
200 nm
(b1)
1 µm

FIG 2

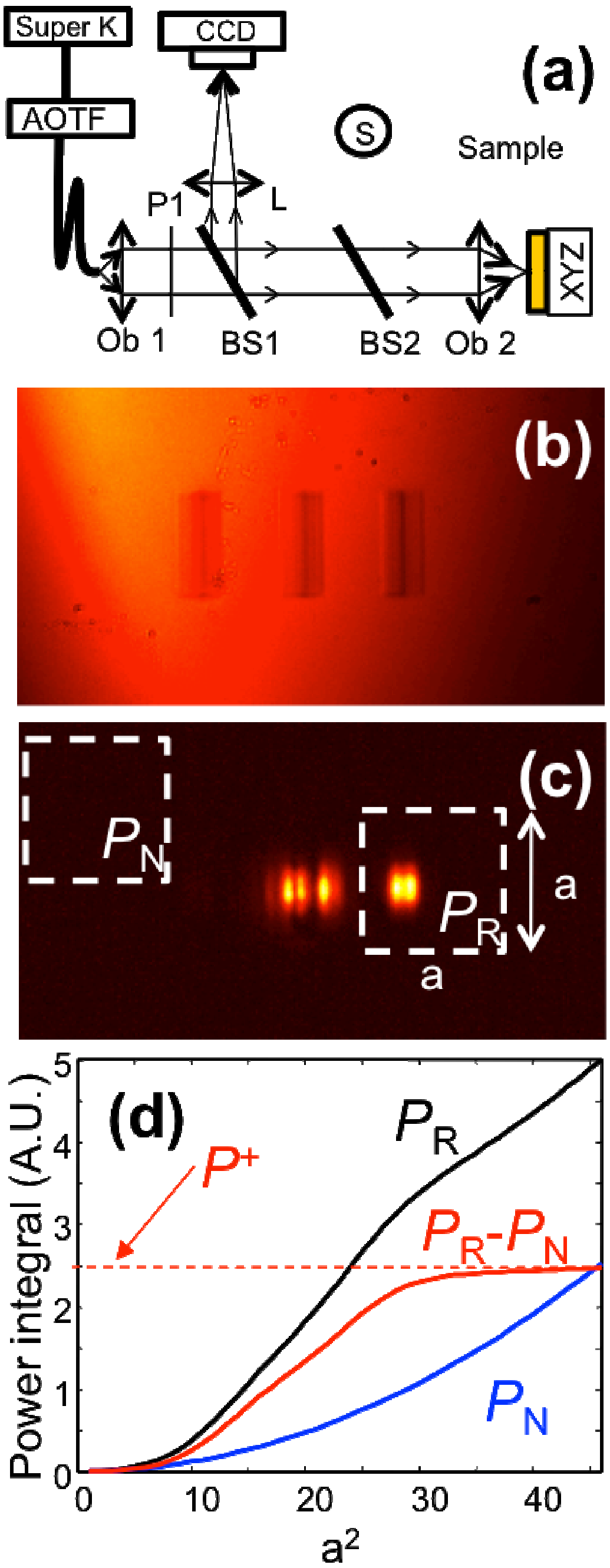
Super K
AOTF
CCD
(a)
S
Sample
P1
L
XYZ
Ob 1
BS1
BS2
Ob 2
(b)
(c)
$P_N$
$P_R$
a
a
(d)
$P_R$
$P^+$
$P_R$-$P_N$
$P_N$
Power integral (A.U.)
$a^2$
0
1
2
3
4
5
0
10
20
30
40

FIG 3

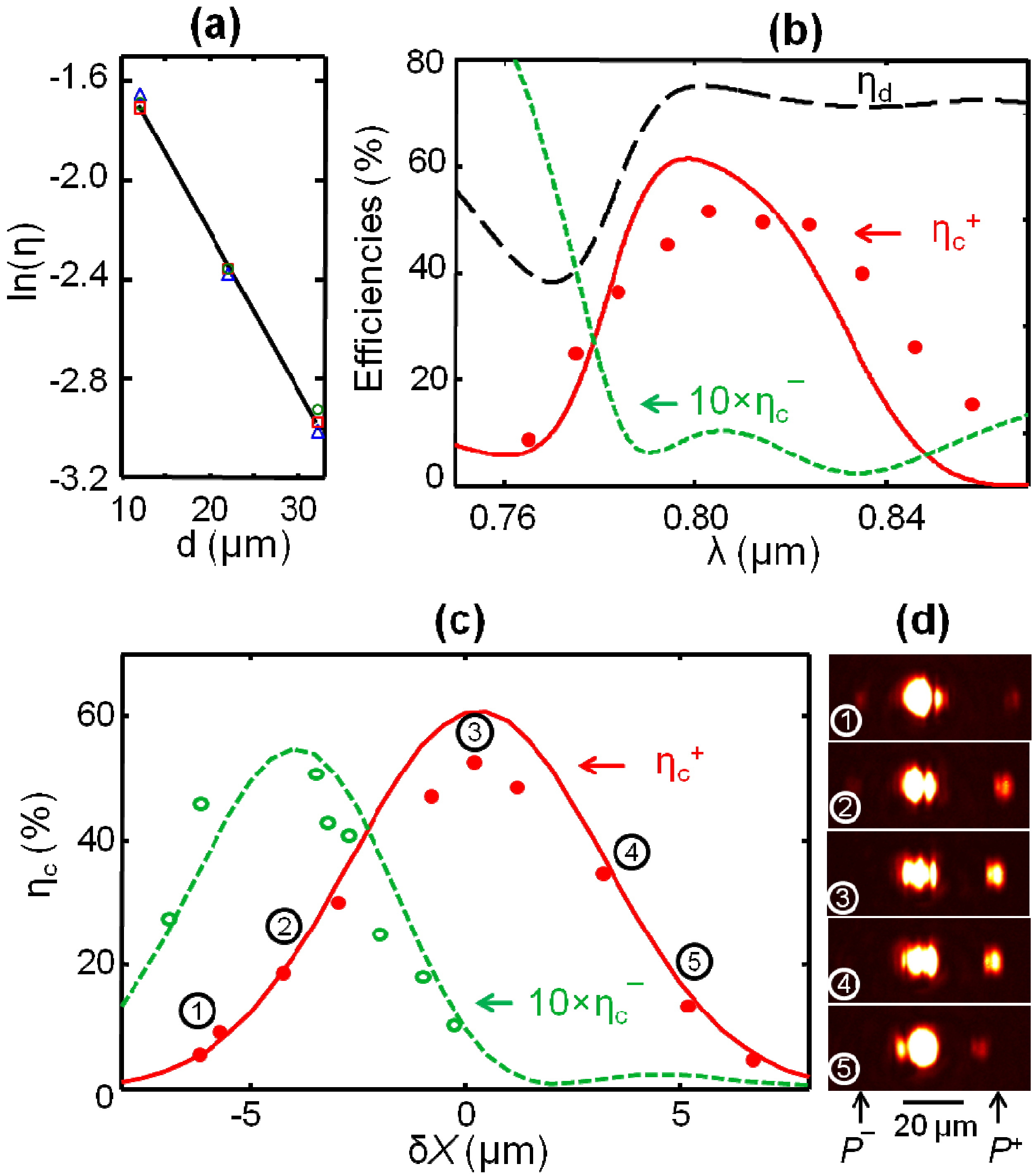
(a)
-1.6
-2.0
-2.4
-2.8
-3.2
ln(η)
10
20
30
d (μm)
(b)
80
60
40
20
0
Efficiencies (%)
0.76
0.80
0.84
λ (μm)
ηd
ηc+
10×ηc−
(c)
60
40
20
0
ηc (%)
-5
0
5
δX (μm)
ηc+
10×ηc−
(d)
20 μm
P−
P+

FIG 4

FIGURE CAPTIONS

**FIG. 1 Designed launcher device. (a)** Interference pattern produced by the incident Gaussian beam ($w_0 = 4.8$ μm) and its reflection on a perfectly flat gold interface at $\lambda = 800$ nm. **(b)** Modulus of the magnetic field $H_z$ scattered by the launcher device illuminated by the normally incident Gaussian beam shown in **(a)**. The incident and the specularly reflected waves have been removed for the sake of clarity. The superimposed curves show the generation strengths, $|\alpha^{+}|^2$ (white) and $|\alpha^{-}|^2$ (green), of the SPP propagating on the right and on the left, respectively. Outside the device ($|x| > x_0$), these two coefficients also represent the right and left launching efficiencies, $\eta_c^{+} = 62\%$ and $\eta_c^{-} = 1\%$, corresponding to a theoretical extinction ratio $E_R \sim 62$. **(c)** Modulus of the magnetic field $H_z$ scattered by the same device, excited by an incident SPP coming from the right side. The launcher now acts as a decoupler, converting the incident SPP into free-space photons. The left inset shows the radiation diagram $R(\theta)$ and the right inset shows the integrated radiated power $P(\theta) = \int_{-\pi/2}^{\theta} R(\theta)d(\theta)$. A large fraction $\eta_d = 75\%$ of the incident SPP energy is radiated out into a narrow cone with a ±10° extraction angle around the surface normal.

**FIG. 2 Sample layout. (a)** Schematic of the sample layout incorporating three SPP launcher devices. The central device, illuminated by a Gaussian beam, launches an SPP on both sides with efficiencies $\eta_c^{+}$ and $\eta_c^{-}$. The devices on both sides are identical to the central launcher and have identical decoupling efficiencies $\eta_d$. $P^{+}$ and $P^{-}$ represent the power of the light radiated into free space on both sides by the decouplers. $P_0/R$ represents the incident beam energy and R is the reflection coefficient for gold. **(b)** SEM pictures of the sample. **(b1)** is an image of the SPP launcher device. **(b2)** is a picture containing the launcher and the two decouplers seperated by a distance *d*. **(b3)** and **(b4)** provide close-ups of a single device, on which gold roughness can be seen on the surface of the film and at the bottom of the grooves.

**FIG. 3 Experimental setup and protocol used to characterize the device. (a) Experimental setup.** It is composed of a supercontinuum laser source (Super K), filtered by an acousto-optic tunable filter (AOTF), two microscope objectives (Ob1 and Ob2), two beam splitters (BS1 and BS2), a CCD camera with 640×480

pixels (9.9 μm)$^2$, a focusing lens (L1: f = 500 mm), a polarizer (P1), a 3-D translation stage (XYZ) and an incoherent light-emitting-device (S) emitting around 670 nm. **Processing the CCD images: (b)** CCD image of the gratings under red incoherent illumination. **(c)** Optical image of the beams scattered by the launcher device. The central spot is the retro-reflection of the incident beam by the central launching device, whereas the right-hand side spot is the light radiated by the SPP-decoupler device. The dashed boxes represent square windows over which we integrate the CCD signal that is proportional to the intensity of the light. **(d)** Integrated radiated powers of image **(c)** as a function of window area $a^2$. The unit of $a$ is in pixels. $P_N$ is the noise power, $P_R$ the power of the raw signal and the asymptotic value of $P_R - P_N$ represents the noise-free power, $P^+$ which does not vary with the area of the integration window for large $a$.

**FIG. 4 Experimental results. (a)** Plot of the natural logarithm of the total coupling/decoupling efficiency $\eta = RP^+/P_0$ as a function of device separation distance $d$. The nine experimental values are plotted as triangles, squares and circles. The linear fit, shown with a solid line, subsequently enables to get an experimental value of the SPP-damping coefficient $l_e$. **(b)** Spectral behavior of the launching efficiencies $\eta_c^+$ and $\eta_c^-$, and the decoupling efficiency $\eta_d$. The solid-red, short dashed-green and long dashed-black curves represent the computational predictions for $\eta_c^+$, $\eta_c^-$ and $\eta_d$ respectively. The corresponding experimental data are shown with the red dots. **(c)** Comparison between the calculated and experimental values for the launching efficiencies $\eta_c^+$ and $\eta_c^-$ as a function of the lateral offset $\delta X$ of the incident beam. The solid-red and dashed-green curves represent the computational predictions for $\eta_c^+$ and $\eta_c^-$. The corresponding experimental data are shown with the red dots and the green-open circles, respectively. **(d)** Optical images acquired with the CCD for several offsets, showing the qualitative variation of the field radiated by the decouplers as the central launching device is laterally shifted relatively to the incident beam.